\documentclass[12pt]{article}
\usepackage{amssymb,amscd,array}
\catcode `\@=11
\@addtoreset{equation}{section}

\newtheorem{thm}{Theorem}[section]

\def\qed{\blacksquare}
\newcommand{\be}{\begin{equation}}
\newcommand{\ee}{\end{equation}}
\newcommand{\bea}{\begin{eqnarray}}
\newcommand{\eea}{\end{eqnarray}}
\newcommand{\R}{\mathbb{R}}
\newcommand{\N}{\mathbb{N}}
\newcommand{\C}{\mathbb{C}}

\begin{document}
\begin{titlepage}

\begin{center}
{\bf \Large{Trivial Lagrangians in the Causal Approach\\}}
\end{center}
\vskip 1.0truecm
\centerline{D. R. Grigore, 
\footnote{e-mail: grigore@theory.nipne.ro}}
\vskip5mm
\centerline{Department of Theoretical Physics, Institute for Physics and Nuclear
Engineering ``Horia Hulubei"}
\centerline{Institute of Atomic Physics}
\centerline{Bucharest-M\u agurele, P. O. Box MG 6, ROM\^ANIA}

\vskip 2cm
\bigskip \nopagebreak
\begin{abstract}
\noindent
We prove the non-uniqueness theorem for the chronological products of a gauge model. We use a cohomological
language where the cochains are chronological products, gauge invariance means a cocycle restriction and
coboundaries are expressions producing zero sandwiched between physical states. Suppose that we
have gauge invariance up to order n of the perturbation theory and we modify the first-order chronological
products by a coboundary (a trivial Lagrangian). Then the chronological products up to order n
get modified by a coboundary also. 
\end{abstract}
\end{titlepage}

\section{Introduction}

The general framework of perturbation theory consists in the construction of 
the chronological products such that Bogoliubov axioms are verified 
\cite{BS}, \cite{EG}, \cite{DF}; for every set of Wick monomials 
$ 
A_{1}(x_{1}),\dots,A_{n}(x_{n}) 
$
acting in some Fock space
$
{\cal H}
$
one associates the operator
$$ 
T(A_{1}(x_{1}),\dots,A_{n}(x_{n})) 
$$ 
which is a distribution-valued operators called chronological product. 

The construction of the chronological products can be done recursively according
to Epstein-Glaser prescription \cite{EG}, \cite{Gl} (which reduces the induction
procedure to a distribution splitting of some distributions with causal support)
or according to Stora prescription \cite{PS} (which reduces the renormalization
procedure to the process of extension of distributions). These products are not
uniquely defined but there are some natural limitation on the arbitrariness. If
the arbitrariness does not grow with $n$ we have a renormalizable theory.An
equivalent point of view uses retarded products \cite{St1}.

Gauge theories describe particles of higher spin. Usually such theories are not
renormalizable. However, one can save renormalizablility using ghost fields.
Such theories are defined in a Fock space
$
{\cal H}
$
with indefinite metric, generated by physical and un-physical fields (called
{\it ghost fields}). One selects the physical states assuming the existence of
an operator $Q$ called {\it gauge charge} which verifies
$
Q^{2} = 0
$
and such that the {\it physical Hilbert space} is by definition
$
{\cal H}_{\rm phys} \equiv Ker(Q)/Im(Q).
$
The space
$
{\cal H}
$
is endowed with a grading (usually called {\it ghost number}) and by
construction the gauge charge is raising the ghost number of a state. Moreover,
the space of Wick monomials in
$
{\cal H}
$
is also endowed with a grading which follows by assigning a ghost number to
every one of the free fields generating
$
{\cal H}.
$
The graded commutator
$
d_{Q}
$
of the gauge charge with any operator $A$ of fixed ghost number
\be
d_{Q}A = [Q,A]
\ee
is raising the ghost number by a unit. It means that
$
d_{Q}
$
is a co-chain operator in the space of Wick polynomials. From now on
$
[\cdot,\cdot]
$
denotes the graded commutator. From
$
Q^{2} = 0
$
one derives
\be
(d_{Q})^{2} = 0.
\label{Q-square}
\ee
 
A gauge theory assumes also that there exists a Wick polynomial of null ghost
number
$
T(x)
$
called {\it the interaction Lagrangian} such that
\be
~[Q, T] = i \partial_{\mu}T^{\mu}
\label{gauge-T}
\ee
for some other Wick polynomials
$
T^{\mu}.
$
This relation means that the expression $T$ leaves invariant the physical
states, at least in the adiabatic limit. Indeed, if this is true we have:
\be
T(f)~{\cal H}_{\rm phys}~\subset~~{\cal H}_{\rm phys}  
\label{phys-inv}
\ee
up to terms which can be made as small as desired (making the test function $f$
flatter and flatter). We call this argument the {\it formal adiabatic limit}.
It is a way to justify from the physical point of view relation (\ref{gauge-T}). Otherwise,
we simply have to postulate it.

In all known models one finds out that there exist a chain
of Wick polynomials
$
T^{\mu},~T^{\mu\nu},\dots
$
such that:
\be
~[Q, T] = i \partial_{\mu}T^{\mu}, \quad
[Q, T^{\mu}] = i \partial_{\nu}T^{\mu\nu}, \quad
[Q, T^{\mu\nu}] = i \partial_{\rho}T^{\mu\nu\rho},\dots
\label{descent}
\ee
It so happens that for all these models the expressions
$
T^{\mu\nu},~T^{\mu\nu\rho},\dots
$
are completely antisymmetric in all indexes; it follows that the chain of
relation stops at the step $4$ (if we work in four dimensions). We can also use
a compact notation
$
T^{I}
$
where $I$ is a collection of indexes
$
I = [\nu_{1},\dots,\nu_{p}]~(p = 0,1,\dots,)
$
and the brackets emphasize the complete antisymmetry in these indexes. All these
polynomials have the same canonical dimension
\be
\omega(T^{I}) = \omega_{0},~\forall I
\ee
and because the ghost number of
$
T \equiv T^{\emptyset}
$
is supposed null, then we also have:
\be
gh(T^{I}) = |I|.
\ee
One can write compactly the relations (\ref{descent}) as follows:
\be
d_{Q}T^{I} = i~\partial_{\mu}T^{I\mu}.
\label{gauge-1}
\ee

For concrete models the equations (\ref{descent}) can stop earlier: for 
instance in the Yang-Mills case we have
$
T^{\mu\nu\rho} = 0
$
and in the case of gravity
$
T^{\mu\nu\rho\sigma} = 0.
$

Now we can construct the chronological products
$
T(T^{I_{1}}(x_{1}),\dots,T^{I_{n}}(x_{n}))
$
according to the recursive procedure. We say that the theory is gauge invariant
in all orders of the perturbation theory if the following set of identities
generalizing (\ref{gauge-1}):
\be
d_{Q}T(T^{I_{1}}(x_{1}),\dots,T^{I_{n}}(x_{n})) = 
i \sum_{l=1}^{n} (-1)^{s_{l}} {\partial\over \partial x^{\mu}_{l}}
T(T^{I_{1}}(x_{1}),\dots,T^{I_{l}\mu}(x_{l}),\dots,T^{I_{n}}(x_{n}))
\label{gauge-n}
\ee
are true for all 
$n \in \N$
and all
$
I_{1}, \dots, I_{n}.
$
Here we have defined
\be
s_{l} \equiv \sum_{j=1}^{l-1} |I|_{j}.
\ee
In particular, the case
$
I_{1} = \dots = I_{n} = \emptyset
$
it is sufficient for the gauge invariance of the scattering matrix, at least
in the adiabatic limit: we have the same argument as for relation (\ref{phys-inv}).

To describe this property in a cohomological framework, we consider that the chronological products are the 
cochains and we define for the operator $\delta$ by 
\be
\delta T(T^{I_{1}}(x_{1}),\dots,T^{I_{n}}(x_{n})) = 
i \sum_{l=1}^{n} (-1)^{s_{l}} {\partial\over \partial x^{\mu}_{l}}
T(T^{I_{1}}(x_{1}),\dots,T^{I_{l}\mu}(x_{l}),\dots,T^{I_{n}}(x_{n})).
\label{der}
\ee
It is easy to prove that we have:
\be
\delta^{2} = 0
\ee
and
\be
[ d_{Q}, \delta ] = 0.
\ee
Next we define 
\be
s \equiv d_{Q} - i \delta
\ee
such that relation (\ref{gauge-n}) can be rewritten as
\be
sT(T^{I_{1}}(x_{1}),\dots,T^{I_{n}}(x_{n})) = 0.
\label{s-gauge-n}
\ee

We note that if we define
\be
\bar{s} \equiv d_{Q} + i \delta
\ee
we have
\be
s\bar{s} = 0, \qquad \bar{s} s = 0
\label{s2}
\ee
so expressions verifying the relation
$
s C = 0
$
can be called {\it cocycles} and expressions of the type
$
\bar{s}B
$
are the {\it coboundaries}. One can build the corresponding cohomology space in the standard way.

The use of this construction is the following. The expressions
$
T^{I}
$
are not unique. Indeed the redefinitions by a coboundary
\be
T^{I} \rightarrow T^{I} + \bar{s}B^{I}
\label{coboundary}
\ee
preserve the relation (\ref{gauge-1}) and (with appropriate restrictions coming from Lorentz invariance
and canonical dimension) it is the most general redefinition preserving gauge invariance  (\ref{gauge-1}).
Expressions of the type
$
\bar{s}B
$
i.e. coboundaries are trivial from the physical point of view: they give zero when restricted to
the physical subspace (in the formal adiabatic limit) so they are {\it trivial Lagrangians}.

We are interested in the following problem. Suppose that we modify the expressions
$
T^{I}
$
by a coboundary (i.e. a trivial Lagrangian). Then in what way would be modified the chronological
products in an arbitrary order $n$? We will prove that if we impose (\ref{gauge-n}) for
$
1, 2,\dots,n
$
the modification of the chronological products in order $n$ is also a coboundary i.e. something trivial
from the physical point of view. This problem was addressed (in the causal formalism) for the 
first time in \cite{D} but no complete proof is provided.

In the next Section we give the essential ingredients for a causal gauge theory. The we will prove the 
result announced above in Section \ref{trivial}.
\newpage

\section{Bogoliubov Axioms}{\label{bogoliubov}}

Suppose that the Wick monomials
$
A_{1},\dots,A_{n}
$
are self-adjoint:
$
A_{j}^{\dagger} = A_{j},~\forall j = 1,\dots,n
$
and of Fermi number
$
f_{i}.
$
We impose the {\it causality} property:
\be
A_{j}(x)~A_{k}(y) = (- 1)^{f_{j}f_{k}}~~A_{k}(y)~A_{j}(x)
\ee
for 
$
(x - y)^{2} < 0
$
i.e.
$
x - y
$
outside the causal cones (this relation is denoted by
$
x \sim y
$).

The chronological products
$ 
T(A_{1}(x_{1}),\dots,A_{n}(x_{n})) \quad n = 1,2,\dots
$
are verifying the following set of axioms:
\begin{itemize}
\item
Skew-symmetry in all arguments
\be
T(\dots,A_{i}(x_{i}),A_{i+1}(x_{i+1}),\dots,) =
(-1)^{f_{i} f_{i+1}} T(\dots,A_{i+1}(x_{i+1}),A_{i}(x_{i}),\dots)
\ee

\item
Poincar\'e invariance: we have a natural action of the Poincar\'e group in the
space of Wick monomials and we impose that for all 
$g \in inSL(2,\C)$
we have:
\be
U_{g} T(A_{1}(x_{1}),\dots,A_{n}(x_{n})) U^{-1}_{g} =
T(g\cdot A_{1}(x_{1}),\dots,g\cdot A_{n}(x_{n}))
\label{invariance}
\ee
where in the right hand side we have the natural action of the Poincar\'e group on
Wick monomials (build from Lorentz covariant free fields).

Sometimes it is possible to supplement this axiom by other invariance
properties: space and/or time inversion, charge conjugation invariance, global
symmetry invariance with respect to some internal symmetry group, supersymmetry,
etc.
\item
Causality: if 
$
x - y 
$
is in the upper causal cone then we denote this relation by
$
x \succeq y
$.
Suppose that we have 
$x_{i} \succeq x_{j}, \quad \forall i \leq k, \quad j \geq k+1$.
then we have the factorization property:
\be
T(A_{1}(x_{1}),\dots,A_{n}(x_{n})) =
T(A_{1}(x_{1}),\dots,A_{k}(x_{k}))~~T(A_{k+1}(x_{k+1}),\dots,A_{n}(x_{n}));
\label{causality}
\ee

\item
Unitarity: We define the {\it anti-chronological products} using a convenient notation introduced
by Epstein-Glaser, adapted to the Grassmann context. If 
$
X = \{j_{1},\dots,j_{s}\} \subset N \equiv \{1,\dots,n\}
$
is an ordered subset, we define
\be
T(X) \equiv T(A_{j_{1}}(x_{j_{1}}),\dots,A_{j_{s}}(x_{j_{s}})).
\ee
Let us consider some Grassmann variables
$
\theta_{j},
$
of parity
$
f_{j},  j = 1,\dots, n
$
and let us define
\be
\theta_{X} \equiv \theta_{j_{1}} \cdots \theta_{j_{s}}.
\ee
Now let
$
(X_{1},\dots,X_{r})
$
be a partition of
$
N = \{1,\dots,n\}
$
where
$
X_{1},\dots,X_{r}
$
are ordered sets. Then we define the sign
$
\epsilon(X_{1},\dots,X_{r})
$
through the relation
\be
\theta_{1} \cdots \theta_{n} = \epsilon(X_{1}, \dots,X_{r})~\theta_{X_{1}} \dots \theta_{X_{r}}
\ee

Then the antichronological products according to
\be
(-1)^{n} \bar{T}(N) \equiv \sum_{r=1}^{n} 
(-1)^{r} \sum_{I_{1},\dots,I_{r} \in Part(N)}
\epsilon(X_{1},\dots,X_{r})~T(X_{1})\cdots T(X_{r})
\label{antichrono}
\ee
Then the unitarity axiom is:
\be
\bar{T}(N) = T(N)^{\dagger}.
\label{unitarity}
\ee
\item
The ``initial condition"
\be
T(A(x)) = A(x).
\ee
\end{itemize}

It can be proved that this system of axioms can be supplemented with
\bea
T(A_{1}(x_{1}),\dots,A_{n}(x_{n}))
\nonumber \\
= \sum \epsilon \quad
<\Omega, T(A^{\prime}_{1}(x_{1}),\dots,A^{\prime}_{n}(x_{n}))\Omega>~~
:A^{\prime\prime}_{1}(x_{1}),\dots,A^{\prime\prime}_{n}(x_{n}):
\label{wick-chrono2}
\eea
where
$A^{\prime}_{i}$
and
$A^{\prime\prime}_{i}$
are Wick submonomials of
$A_{i}$
such that
$A_{i} = :A^{\prime}_{i} A^{\prime\prime}_{i}:$
and the sign
$\epsilon$
takes care of the permutation of the Fermi fields; here
$\Omega$
is the vacuum state. This is called the {\it Wick expansion property}. 

We can also include in the induction hypothesis a limitation on the order of
singularity of the vacuum averages of the chronological products associated to
arbitrary Wick monomials
$A_{1},\dots,A_{n}$;
explicitly:
\be
\omega(<\Omega, T^{A_{1},\dots,A_{n}}(X)\Omega>) \leq
\sum_{l=1}^{n} \omega(A_{l}) - 4(n-1)
\label{power}
\ee
where by
$\omega(d)$
we mean the order of singularity of the (numerical) distribution $d$ and by
$\omega(A)$
we mean the canonical dimension of the Wick monomial $W$.

Up to now, we have defined the chronological products only for self-adjoint Wick monomials 
$
W_{1},\dots,W_{n}
$
but we can extend the definition for Wick polynomials by linearity.

The construction of Epstein-Glaser is based on the following recursive procedure. Suppose that
we know the chronological products up to order $n - 1$. Then we define the following expression:
\be
D(N) \equiv - \sum_{(X,Y) \in Part(N)}~( - 1)^{|Y|}~\epsilon(X,Y)~[ \bar{T}(X), T(Y) ]
\label{D-n}
\ee
where the partitions
$
(X,Y)
$
are restricted by
$
n \in X, Y \not= \emptyset,
$
$
|Y|
$
is the cardinal of $Y$ and the commutator is graded. These restrictions guarantee that
$
|X|, |Y| < n
$
so the expressions in the right-hand side of the previous expression are known by the induction
hypothesis. Then it can be proved that the expression
$
D(N) = D(A_{1}(x_{1}),\dots,A_{n}(x_{n}))
$
has causal support in the variables
$
x_{1} - x_{n},\dots,x_{n-1} - x_{n}
$
; accordingly is called the {\it causal commutator}. One can causally split 
$
D(N)
$
as
\be
D(N) = D^{\rm adv}(N) - D^{\rm ret}(N)
\ee
with $D^{\rm adv}(N)$ (resp. $D^{\rm ret}(N)$) with support in the upper (resp. lower) light cone.
From these expression one can construct the chronological products
$
T(N)
$
in order $n$ in a standard way.
\newpage
\section{Trivial Lagrangians\label{trivial}}

Here we have proved the following
\begin{thm}
Suppose the chronological products are chosen such that we have gauge invariance (\ref{gauge-n}) up to order $n$
and we modify the first order chronological products (the interaction Lagrangian) by a coboundary (a trivial
Lagrangian):
\be
T^{I} \rightarrow T^{I} + T^{I}_{0}
\label{redefinition}
\ee
where 
\be
T^{I}_{0}\equiv \bar{s}B^{I}
\label{sB}
\ee
is a coboundary. Then the chronological products, up to order $n$ get modified by a coboundary also:
\be
T(T^{I_{1}}(x_{1}),\dots,T^{I_{n}}(x_{n})) \rightarrow T(T^{I_{1}}(x_{1}),\dots,T^{I_{n}}(x_{n})) 
+ T_{0}(T^{I_{1}}(x_{1}),\dots,T^{I_{n}}(x_{n})) 
\ee
where the expression 
$
T_{0}(T^{I_{1}}(x_{1}),\dots,T^{I_{n}}(x_{n})) 
$
is a coboundary.
\end{thm}

{\bf Proof:} Let us work first in the {\bf second order} of the perturbation theory. From (\ref{D-n}) we see that the second-order
causal commutator coincides with the usual commutator
\be
D(A(x), B(y)) = [ A(x), B(y) ].
\label{D2}
\ee

Suppose we make the redefinition (\ref{redefinition}); then we have
\be
T(T^{I}(x),T^{J}(y)) \rightarrow T(T^{I}(x),T^{J}(y)) + T_{0}(T^{I}(x),T^{J}(y)) 
\ee
where
\be
T_{0}(T^{I}(x),T^{J}(y)) = T(T_{0}^{I}(x),T^{J}(y)) + T(T^{I}(x),T_{0}^{J}(y))
+ T(T_{0}^{I}(x),T_{0}^{J}(y)).
\label{T0-2}
\ee
We easily determine by direct computations that
\be
D(T_{0}^{I}(x),T^{J}(y)) + D(T^{I}(x),T_{0}^{J}(y)) = \bar{s}B_{0}(T^{I}(x),T^{J}(y)) 
\label{sD-2}
\ee
where
\be
B_{0}(T^{I}(x),T^{J}(y))  \equiv D(B^{I}(x),T^{J}(y)) + (- 1)^{|I|}~D(T^{I}(x),B^{J}(y)) 
\label{B-2}
\ee
To see how this works let us compute
\bea
D(T_{0}^{I}(x),T^{J}(y)) = D(\bar{s}B^{I}(x),T^{J}(y)) = [ \bar{s}B^{I}(x),T^{J}(y)] 
\nonumber \\
= [ d_{Q}B^{I}(x) + i \partial_{\mu}B^{I\mu}(x), T^{J}(y)] 
\nonumber \\
= d_{Q}~[ B^{I}(x), T^{J}(y) ] + ~(-1)^{|I|}~[ B^{I}(x), d_{Q}T^{J}(y) ]
+ i {\partial \over \partial x^{\mu}} [ B^{I\mu}(x), T^{J}(y) ]
\label{step1}
\eea
where we have used the fact that
$
d_{Q}
$
verifies the (graded) Leibniz rule. In the second term above we use first-order gauge
invariance (\ref{gauge-1}) and obtain
\bea
D(T_{0}^{I}(x),T^{J}(y)) = 
\nonumber \\
= d_{Q}~[ B^{I}(x), T^{J}(y) ] + ~i~(-1)^{|I|}~{\partial \over \partial y^{\mu}}~[ B^{I}(x), T^{J}(y) ]
+ i {\partial \over \partial x^{\mu}} [ B^{I\mu}(x), T^{J}(y) ].
\label{step2}
\eea
The second term of left hand side of (\ref{sD-2}) is computed in the same way and regrouping the terms we get the 
result.

Now we see that in (\ref{sD-2}) both sides have causal support, so the causal splitting
produces
\be
T(T_{0}^{I}(x),T^{J}(y)) + T(T^{I}(x),T_{0}^{J}(y)) = \bar{s}B^{F}_{0}(T^{I}(x),T^{J}(y)) 
\label{sT-2}
\ee
where
\be
B^{F}_{0}(T^{I}(x),T^{J}(y))  \equiv T(B^{I}(x),T^{J}(y)) + (- 1)^{|I|}~T(T^{I}(x),B^{J}(y)) 
\label{BF-2}
\ee
This means that the first two terms from the right-hand side of (\ref{T0-2}) are a coboundary.
Because we have
$
sT^{I}_{0} = s\bar{s}B^{I} = 0
$
according to (\ref{s2}) it follows that we can repeat the computations leading to (\ref{sD-2}) + (\ref{B-2})
with
$
T^{I} \rightarrow T_{0}^{I}
$
and we obtain instead of (\ref{sT-2})
\be
T(T_{0}^{I}(x),T_{0}^{J}(y)) = {1 \over 2}~\bar{s}B^{F}_{0}(T_{0}^{I}(x), T_{0}^{J}(y)) 
\label{sT0-3}
\ee
so the last term of (\ref{T0-2}) is a coboundary. In conclusion we have the desired property in the 
second-order of perturbation theory:
\be
T_{0}(T^{I}(x),T^{J}(y)) = \bar{s}B^{F}(T^{I}(x),T^{J}(y)) 
\ee
where
\be
B^{F}(T^{I}(x),T^{J}(y)) = B^{F}_{0}(T^{I}(x),T^{J}(y)) + {1 \over 2}~B^{F}_{0}(T_{0}^{I}(x), T_{0}^{J}(y)). 
\ee

We have proved that if we modify the interaction Lagrangian
$
T^{I}
$
by a trivial Lagrangian (a coboundary), then the second order chronological products get modified 
by a coboundary also.
\newpage
(ii) It is illuminating to push the proof to the {\bf third} order of the perturbation theory. We suppose that
we have fixed the second-order chronological products such that we have gauge invariance in the 
second-order (\ref{gauge-n}) for 
$
n = 2.
$
From (\ref{D-n}) we have similarly with (\ref{D2}):
\bea
D(A(x),B(y),C(z)) = - [\bar{T}(A(x),B(y)),C(z)] 
\nonumber\\
- (-1)^{|B||C|}~[T(A(x),C(z)), B(y)] - (-1)^{|A|(|B|+|C|)}~[T(B(y),C(z)), A(x)].
\label{D3}
\eea
Also, similarly to (\ref{T0-2}), we have
\bea
T_{0}(T^{I}(x),T^{J}(y),T^{K}(z)) =
\nonumber \\
T(T_{0}^{I}(x),T^{J}(y),T^{K}(z)) + T(T^{I}(x),T_{0}^{J}(y),T^{K}(z)) + T(T^{I}(x),T^{J}(y),T_{0}^{K}(z)) 
\nonumber \\
+ T(T^{I}(x),T_{0}^{J}(y),T_{0}^{K}(z)) + T(T_{0}^{I}(x),T^{J}(y),T_{0}^{K}(z)) + T(T_{0}^{I}(x),T_{0}^{J}(y),T^{K}(z)) 
\nonumber\\
+ T_{0}(T_{0}^{I}(x),T_{0}^{J}(y),T_{0}^{K}(z)).
\label{T0-3}
\eea
Guided by the previous (second-order) analysis we prove by direct computation that
\bea
D(T_{0}^{I}(x),T^{J}(y),T^{K}(z)) + D(T^{I}(x),T_{0}^{J}(y),T^{K}(z)) + D(T^{I}(x),T^{J}(y),T_{0}^{K}(z)) 
\nonumber \\
= \bar{s}B_{0}(T^{I}(x),T^{J}(y),T^{K}(z))
\label{sD-3}
\eea
where
\bea
B_{0}(T^{I}(x),T^{J}(y),T^{K}(z)) = D(B^{I}(x),T^{J}(y),T^{K}(z)) 
\nonumber \\
+ (- 1)^{|I|}~D(T^{I}(x),B^{J}(y),T^{K}(z)) + 
(- 1)^{|I| + |J|}~D(T^{I}(x),T^{J}(y),B^{K}(z)). 
\label{B3}
\eea
In this proof gauge invariance in the second-order must be used as in (\ref{step1}) 
$
\Rightarrow
$
(\ref{step2}) above. Now both hand sides of (\ref{sD-3}) are with causal support, so the causal
splitting gives
\bea
T(T_{0}^{I}(x),T^{J}(y),T^{K}(z)) + T(T^{I}(x),T_{0}^{J}(y),T^{K}(z)) + T(T^{I}(x),T^{J}(y),T_{0}^{K}(z)) 
\nonumber \\
= \bar{s}B^{F}_{0}(T^{I}(x),T^{J}(y),T^{K}(z))
\label{sT-3}
\eea
where
\bea
B^{F}_{0}(T^{I}(x),T^{J}(y),T^{K}(z)) = T(B^{I}(x),T^{J}(y),T^{K}(z)) 
\nonumber \\
+ (- 1)^{|I|}~T(T^{I}(x),B^{J}(y),T^{K}(z)) + 
(- 1)^{|I| + |J|}~T(T^{I}(x),T^{J}(y),B^{K}(z)).
\label{BF-3}
\eea
The last two terms of (\ref{T0-3}) can be easily computed: Using gauge invariance in the second order and 
(\ref{sT-2}) + (\ref{sT0-3}) we see that the expression
$$
T(T^{I}(x) + \alpha T_{0}^{I}(x),T^{J}(y) + \alpha T^{J}_{0}(y))
$$
is gauge invariant for an arbitrary 
$
\alpha \in \R
$
so the previous proof of (\ref{sT-3}) + (\ref{BF-3}) with stays true for 
$
T^{I} \rightarrow T^{I} + \alpha T^{I}_{0}.
$
The coefficients of 
$
\alpha^{2}
$
and
$
\alpha^{3}
$
give the coboundary structure of the last two terms of (\ref{T0-3}) and we have the result for 
$
n = 3.
$

\newpage
(iii) Finally we go to the general case of arbitrary $n$. We want to determine the expression
\bea
T_{0}(T^{I_{1}}(x_{1}),\dots,T^{I_{n}}(x_{n})) \equiv
\nonumber\\
T(T^{I_{1}}(x_{1}) + T_{0}^{I_{1}}(x_{1}),\dots,T^{I_{n}}(x_{n}) + T_{0}^{I_{n}}(x_{n}))
- T(T^{I_{1}}(x_{1},\dots,T^{I_{n}}(x_{n}))
\label{T0-n}
\eea
and prove that it is a coboundary. We introduce the following notations:
$$
T_{l_{1},\dots,l_{r}}(T^{I_{1}}(x_{1}),\dots,T^{I_{n}}(x_{n}))
$$
is obtained from 
$$
T(T^{I_{1}}(x_{1}),\dots,T^{I_{n}}(x_{n}))
$$
by the substitutions
$$
T^{I_{l_{1}}} \rightarrow T_{0}^{I_{l_{1}}}, \dots, T^{I_{l_{r}}} \rightarrow T_{0}^{I_{l_{r}}}
$$
so it easily follows that
\be
T_{0}(T^{I_{1}}(x_{1}),\dots,T^{I_{n}}(x_{n}))
= \sum_{r=1}^{n}~\sum_{l_{1} < \cdots < l_{r}}~T_{l_{1},\dots,l_{r}}(T^{I_{1}}(x_{1}),\dots,T^{I_{n}}(x_{n})).
\label{T0}
\ee
We will prove that all the expressions
$$
T_{l_{1},\dots,l_{r}}(T^{I_{1}}(x_{1}),\dots,T^{I_{n}}(x_{n}))
$$
are coboundaries. As in the cases 
$
n = 2, 3
$
from above, we first prove a generalization of (\ref{sT-2}) + (\ref{BF-2}) and (\ref{sT-3}) + (\ref{BF-3}) by induction.
More precisely, we suppose that we have fixed gauge invariance (\ref{gauge-n}) up to the order 
$
n - 1
$
and proved
\be
\sum_{l=1}^{p}~T(T^{I_{1}}(x_{1}),\dots,T_{0}^{I_{l}}(x_{l}),\dots,T^{I_{n}}(x_{n})) =
\bar{s}B^{F}_{0}(T^{I_{1}}(x_{1}),\dots,T^{I_{n}}(x_{n}))
\label{sT-p}
\ee
where
\be
B^{F}_{0}(T^{I_{1}}(x_{1}),\dots,T^{I_{n}}(x_{n})) = \sum_{l=1}^{p}~\prod_{j < l}~(-1)^{f_{j}}~
T(T^{I_{1}}(x_{1}),\dots,B^{I_{l}}(x_{l}),\dots,T^{I_{n}}(x_{n}))
\label{BF-n}
\ee
for
$
p = 1, \dots, n - 1.
$
We want to prove the same result for 
$
p = n.
$
We determine the sum of causal commutators
\be
D_{n} \equiv \sum_{l=1}^{p}~D(T^{I_{1}}(x_{1}),\dots,T_{0}^{I_{l}}(x_{l}),\dots,T^{I_{n}}(x_{n}))
\label{Dn}
\ee
using the definition (\ref{D-n}) with the substitution 
$
T^{I_{l}} \rightarrow T_{0}^{I_{l}};
$
we have two type of terms: with 
$
l \in X
$
and with
$
l \in Y
$
\bea
D_{n}  = - \sum_{l=1}^{p}~
(\sum_{(X,Y) \in Part(N), l \in X}~(-1)^{|Y|}~\epsilon(X,Y)~[ \bar{T}_{sB_{l}}(X), T(Y) ]
\nonumber\\
+ \sum_{(X,Y) \in Part(N), l \in Y}~(-1)^{|Y|}~\epsilon(X,Y)~[ \bar{T}(X), T_{sB_{l}}(Y) ])
\label{D1}
\eea
where the expression
$
\bar{T}_{sB_{l}}(X), T_{sB_{l}}(Y)
$
are obtained from
$
\bar{T}(X), T(Y)
$
with the substitution 
$
T^{I_{l}} \rightarrow \bar{s}B^{I_{l}}.
$
We invert the order of summation and obtain
\bea
D_{n}  = - \sum_{(X,Y) \in Part(N)}~(-1)^{|Y|}~\epsilon(X,Y)~[ \sum_{l \in X}~\bar{T}_{sB_{l}}(X), T(Y) ]
\nonumber\\
- \sum_{(X,Y) \in Part(N)}~(-1)^{|Y|}~\epsilon(X,Y)~[ \bar{T}(X), \sum_{l \in Y}~T_{sB_{l}}(Y) ].
\label{Dn2}
\eea
Now, the restrictions
$
n \in X, Y \not= \emptyset
$
from the definition of the causal commutator implies that
$
|X|, |Y| < n
$
so we can apply the induction hypothesis and get
\bea
D_{n}  = - \sum_{(X,Y) \in Part(N)}~(-1)^{|Y|}~\epsilon(X,Y)~[ \bar{s}{\bar{B}}^{F}_{0}(X), T(Y) ]
\nonumber\\
- \sum_{(X,Y) \in Part(N)}~(-1)^{|Y|}~\epsilon(X,Y)~[ \bar{T}(X), \bar{s}B^{F}_{0}(Y) ].
\label{Dn3}
\eea
We compute the two commutators as before; for instance
\bea
[ \bar{s}\bar{B}^{F}_{0}(X), T(Y) ] = [ d_{Q}{\bar{B}}^{F}_{0}(X) + i~\delta_{X}\bar{B}^{F}_{0}(X), T(Y) ]
\nonumber\\
= d_{Q}[ \bar{B}^{F}_{0}(X), T(Y) ] + (-1)^{\phi_{X}}~[ \bar{B}^{F}_{0}(X), d_{Q}T(Y) ]
+ i~\delta_{X}[ \bar{B}^{F}_{0}(X), T(Y) ] 
\nonumber
\eea
where the operator
$
\delta_{X}
$
is the operator (\ref{der}) applied to a cocycle depending only on the variables
$
x_{j},~j \in X
$
and
\be
\phi_{X} \equiv \sum_{j \in X}~f_{j}
\ee
is the Fermi number of 
$
T(X).
$
Now we apply the gauge invariance induction hypothesis to express 
$
d_{Q}T(Y) 
$
as
$
i~\delta_{Y}T(Y)
$
and finally
\be
[ \bar{s}{\bar{B}}^{F}_{0}(X), T(Y) ] = \bar{s} [\bar{B}^{F}_{0}(X), T(Y) ].
\ee
We do the same type of computation for the second commutator from (\ref{Dn3}) and end up with
\be
D_{n} = \bar{s}B
\label{Dn4}
\ee
where
\bea
B = - \sum_{(X,Y) \in Part(N)}~(-1)^{|Y|}~\epsilon(X,Y)~[ \bar{B}^{F}_{0}(X), T(Y) ]
\nonumber\\
- \sum_{(X,Y) \in Part(N)}~(-1)^{|Y|}~(-1)^{\phi_{X}}~\epsilon(X,Y)~[ \bar{T}(X), B^{F}_{0}(Y) ].
\label{B}
\eea

The previous expression can be rewritten using the induction hypothesis under the form
\be
B^{F}_{0}(X) = \sum_{l \in X}~(-1)^{\phi_{X,l}}~T_{B_{l}}(X),\qquad |X| < n
\ee
where
\be
\phi_{X,l} \equiv \sum_{j \in X, l < l}~f_{j}
\ee
(remember that $X$ is an ordered set) and
$
T_{B_{l}}(X)
$
is obtained from
$
T(X)
$
with the substitution 
$
T^{I_{l}} \rightarrow B^{I_{l}};
$
a similar formula is true for 
$
\bar{B}^{F}_{0}.
$

We substitute in (\ref{B}), invert the order of summation and dealing carefully with the signs we obtain
\be
B = \sum_{l=1}^{n}~\prod_{j < l}~(-1)^{f_{j}}~
D(T^{I_{1}}(x_{1},\dots,B^{I_{l}}(x_{l}),\dots,T^{I_{n}}(x_{n})).
\ee
It follows that in (\ref{Dn4}) both sides are causal expressions, so the causal splitting gives (\ref{sT-p}) + (\ref{BF-n}) for 
$
p = n.
$

To prove that the expression
$
T_{0}
$
from (\ref{T0-n}) is a coboundary we proceed as follows. We replace the induction hypothesis
(\ref{sT-p}) + (\ref{BF-n}) by a stronger induction hypothesis, namely we suppose that we have for 
$
p = 1,\dots, n -1
$
and
$
r < p
$
the following relation
\be
\sum_{l_{1} <\dots <l_{r}}~T_{l_{1},\dots,l_{r}}(T^{I_{1}}(x_{1}),\dots,T^{I_{n}}(x_{n})) =
\bar{s}B^{F}_{r-1}(T^{I_{1}}(x_{1}),\dots,T^{I_{n}}(x_{n}))
\label{sT-pr}
\ee
where
\bea
B^{F}_{r}(T^{I_{1}}(x_{1}),\dots,T^{I_{n}}(x_{n})) = {1 \over r+1}~\sum_{s=1}^{p}~\prod_{j < s}~(-1)^{f_{j}}~\times
\nonumber\\
\times \sum_{l_{1} <\dots <l_{r}}~ 
T_{l_{1},\dots,l_{r}}(T^{I_{1}}(x_{1}),\dots,B^{I_{l}}(x_{l}),\dots,T^{I_{n}}(x_{n}))
\label{BF-nr}
\eea
where in the sum over
$
l_{1} <\dots <l_{r}
$
we impose
$
\{l_{1},\dots,l_{r}\} \cap \{s\} = \emptyset.
$

In this case we can easily prove that the expressions
$$
T(T^{I_{1}}(x_{1}) + \alpha~T_{0}^{I_{1}}(x_{1}),\dots,T^{I_{n}}(x_{n}) + \alpha~T_{0}^{I_{p}}(x_{p})), ~p < n
$$
are gauge invariant in the sense (\ref{gauge-n}) so we can reconsider the proof of (\ref{sT-p}) + (\ref{BF-n}) for
$
p = n
$
with
$
T^{I} \rightarrow T^{I} + \alpha~T_{0}^{I}
$
for an arbitrary
$
\alpha \in \R. 
$
All expressions are polynomials in $\alpha$ and the coefficient of
$
\alpha^{r}
$
gives exactly the relations (\ref{sT-pr}) + (\ref{BF-nr}) 
and this finishes the induction. Now we have from (\ref{T0})
\be
T_{0}(T^{I_{1}}(x_{1}),\dots,T^{I_{n}}(x_{n})) =
\bar{s}B^{F}(T^{I_{1}}(x_{1}),\dots,T^{I_{n}}(x_{n}))
\ee
i.e. a coboundary, where
\be
B^{F} = \sum_{r=0}^{n-1}~B^{F}_{r}
\ee
and this finishes the proof.
$\qed$


\newpage


\begin{thebibliography}{99}

\bibitem{BS}
N. N. Bogoliubov, D. Shirkov,
``{\it Introduction to the Theory of Quantized Fields}",
John Wiley and Sons, 1976 (3rd edition)

\bibitem{DF}
M. D\"utsch, K. Fredenhagen,
``{\it A Local (Perturbative) Construction of Observables in Gauge Theories:
the Example of QED}",\\
hep-th/9807078, Commun. Math. Phys. {\bf 203} (1999) 71-105

\bibitem{D}
M. D\"utsch,
``{\it Non-uniqueness of quantized Yang–Mills theories}",\\
hep-th/9606100, J. Phys. A: Math. Gen. {\bf 29} (1996) 7597–7617

\bibitem{EG}
H. Epstein, V. Glaser,
``{\it The R\^ole of Locality in Perturbation Theory}",\\
Ann. Inst. H. Poincar\'e {\bf 19 A} (1973) 211-295

\bibitem{Gl}
V. Glaser,
``{\it Electrodynamique Quantique}",
L'enseignement du 3e cycle de la physique en Suisse Romande (CICP), Semestre
d'hiver 1972/73

\bibitem{YM} D. R. Grigore
``{\it On the Uniqueness of the Non-Abelian Gauge Theories in Epstein-Glaser 
Approach to Renormalisation Theory}", \\
hep-th/9806244, Romanian J. Phys. {\bf 44} (1999) 853-913

\bibitem{PS}
G. Popineau, R. Stora, 
``{\it A Pedagogical Remark on the Main Theorem of Perturbative Renormalization
Theory}", unpublished preprint

\bibitem{Sc1}
G. Scharf,
``{\it Finite Quantum Electrodynamics: The Causal Approach}",
(second edition) Springer, 1995

\bibitem{Sc2}
G. Scharf,
``{\it Quantum Gauge Theories. A True Ghost Story}",
John Wiley, 2001
and ``{\it Quantum Gauge Theories - Spin One and Two}",
Google books, 2010

\bibitem{Sto1}
R. Stora,
``{\it Lagrangian Field Theory}",
Les Houches lectures, Gordon and Breach, N.Y., 1971, 
C. De Witt, C. Itzykson eds.

\bibitem{St1}
O. Steinmann,
``{\it Perturbation Expansions in Axiomatic Field Theory}",
Lect. Notes in Phys. {\bf 11}, Springer, 1971

\end{thebibliography}
\end{document}